\documentclass[]{spie}
\usepackage{import}
\usepackage{graphicx}
\usepackage{authblk}
\usepackage{acro}
\usepackage{subcaption}
\usepackage[notransparent]{svg}
\title{Uncertainty aware and explainable diagnosis of retinal disease}
\author[a,b]{Amitojdeep Singh}
\author[a,b]{Sourya Sengupta}
\author[a]{Mohammed Abdul Rasheed}
\author[a]{Varadharajan Jayakumar}
\author[a,b]{Vasudevan Lakshminarayanan}

\affil[a]{Theoretical and Experimental Epistemology Laboratory, School of Optometry and Vision Science, University of Waterloo, Waterloo, ON, Canada}
\affil[b]{Department of Systems Design Engineering, University of Waterloo, Waterloo, ON, Canada}

\authorinfo{E-mail: amitojdeep.singh@uwaterloo.ca}

\date{}
\DeclareAcronym{cnn}{
  short = CNN ,
  long  = convolutional neural network
}
\DeclareAcronym{mse}{
  short = MSE ,
  long  = mean squared error
}
\DeclareAcronym{gbp}{
  short = GBP ,
  long  = guided backpropagation
}
\DeclareAcronym{deeplift}{
  short = DeepLIFT ,
  long  = Deep Learning Important FeaTures
}
\DeclareAcronym{shap}{
  short = SHAP ,
  long  = SHapley Additive exPlanations
}
\DeclareAcronym{ig}{
  short = IG ,
  long  = integrated gradients
}
\DeclareAcronym{lrp}{
  short = LRP ,
  long  = layer wise relevance propagation
}
\DeclareAcronym{svm}{
  short = SVM ,
  long  = support vector machines
}
\DeclareAcronym{dr}{
  short = DR,
  long  = diabetic retinopathy
}
\DeclareAcronym{eg}{
  short = EG,
  long  = expressive gradients
}
\DeclareAcronym{cnv}{
  short =  CNV,
  long  =  choroidal neovascularization
}
\DeclareAcronym{dme}{
  short =  DME,
  long  = diabetic macular edema
}
\DeclareAcronym{amd}{
  short = AMD,
  long  = age related macular degeneration
}
\DeclareAcronym{fda}{
  short = FDA,
  long  = Food and Drug Administration
}
\DeclareAcronym{xai}{
  short = XAI ,
  long  = explainable AI
}
\DeclareAcronym{ai}{
  short = AI,
  long  =  artificial intelligence
}
\DeclareAcronym{rmse}{
  short = RMSE ,
  long  = root mean squared error
}
\DeclareAcronym{gdpr}{
  short = GDPR,
  long  =  General Data Protection Regulation
}
\DeclareAcronym{relu}{
  short = ReLU ,
  long  = rectified linear unit
}
\DeclareAcronym{oct}{
  short = OCT ,
  long  = optical coherence tomography
}

\DeclareAcronym{cad}{
  short = CAD ,
  long  = computer-aided diagnostic
}
\DeclareAcronym{gpu}{
  short = GPU ,
  long  = graphics processing units
}
\DeclareAcronym{spectral domain oct}{
  short =  SD-OCT,
  long  = spectral domain OCT
}
\DeclareAcronym{frequency domain oct}{
  short =  FD-OCT,
  long  = frequency domain OCT
}
\DeclareAcronym{time domain oct}{
  short =  TD-OCT,
  long  = time domain OCT
}
\DeclareAcronym{polarization sensitive oct}{
  short =  PS-OCT,
  long  = polarization sensitive OCT
}
\DeclareAcronym{swept source oct}{
  short =  SS-OCT,
  long  = swept source OCT
}
\DeclareAcronym{onh}{
  short =  ONH,
  long  = optic nerve head
}
\DeclareAcronym{ivus}{
  short =  IVUS,
  long  = intravascular ultrasound
}
\DeclareAcronym{aoslo}{
  short =  AOSLO,
  long  = adaptive optics scanning laser/light ophthalmoscopy
}
\DeclareAcronym{ma}{
  short =  MA,
  long  = microaneurysms
}
\DeclareAcronym{ex}{
  short =  EX,
  long  = exudates
}
\DeclareAcronym{he}{
  short =  HE,
  long  = haemorrhages
}
\DeclareAcronym{elm}{
  short =  ELM,
  long  = external limiting membrane
}
\DeclareAcronym{etdrs}{
  short =  ETDRS,
  long  =  Early treatment diabetic retinopathy study
}
\DeclareAcronym{lbp}{
  short =  LBP,
  long  =  linear binary pattern
}
\DeclareAcronym{hog}{
  short =  HOG,
  long  =  histogram of gradient
}
\DeclareAcronym{glcm}{
  short =  GLCM,
  long  =  gray-level co-occurrence matrix
}

\DeclareAcronym{rf}{
  short =  RF,
  long  =  random forest
}
\DeclareAcronym{ann}{
  short =  ANN,
  long  =  artificial neural network
}
\DeclareAcronym{pca}{
  short =  PCA,
  long  =  principle component analysis
}
\DeclareAcronym{auc}{
  short =  AUC,
  long  =  area under curve
}
\DeclareAcronym{acc}{
  short =  Acc,
  long  =  accuracy
}
\DeclareAcronym{sn}{
  short =  SN,
  long  =  sensitivity
}
\DeclareAcronym{sp}{
  short =  SP,
  long  =  specificity
}
\DeclareAcronym{nw}{
  short =  NW,
  long  =  Koversi non-orthogonal wavelet 
}
\DeclareAcronym{lpnd}{
  short =  LPND,
  long  =  Laplacian pyramid nonlinear diffusion
}
\DeclareAcronym{mlc}{
  short =  MLC,
  long  =  machine learning classifiers
}
\DeclareAcronym{mlp}{
  short =  MLP,
  long  =  multilayer perceptron
}
\DeclareAcronym{rnfl}{
  short =  RNFL,
  long  =  retinal nerve fibre layer
}

\DeclareAcronym{fcn}{
  short =  FCN,
  long  =  fully convolutional neural network
}
\DeclareAcronym{me}{
  short =  ME,
  long  =  mean error
}
\DeclareAcronym{mape}{
  short =  MAPE,
  long  =  mean absolute percentage error
}
\DeclareAcronym{pr}{
  short =  PR,
  long  =  precision
}
\DeclareAcronym{sd}{
  short = SD,
  long  =  standard deviation
}
\DeclareAcronym{vgg}{
  short =  VGG,
  long  =  Visual geometry group
}

\DeclareAcronym{mgrf}{
  short =  MGRF,
  long  =  Markov Gibbs random field
}
\DeclareAcronym{gan}{
  short =  GAN,
  long  =  generative adversarial network
}
\DeclareAcronym{od}{
  short =  OD,
  long  =  optic disc
}
\DeclareAcronym{oc}{
  short =  OC,
  long  =  optic cup
}
\DeclareAcronym{knn}{
 short  =  kNN,
 long   =  k nearest neighbor
}
\DeclareAcronym{ga}{
  short =  GA,
  long  =  geographic atrophy
}
\DeclareAcronym{lstm}{
  short =  LSTM,
  long  =  long short term memory
}
\DeclareAcronym{ssim}{
  short =  SSIM,
  long  =  structural similarity index measurement
}
\DeclareAcronym{csr}{
    short = CSR,
    long = central serous retinopathy
}
\DeclareAcronym{irf}{
    short = IRF,
    long = intra-retinal fluid
}
\DeclareAcronym{srf}{
    short = SRF,
    long = sub-retinal fluid
}
\DeclareAcronym{ped}{
    short = PED,
    long = pigment epithelial detachment
}
\DeclareAcronym{clahe}{
    short = CLAHE,
    long = contrast limited adaptive histogram equalization
}
\DeclareAcronym{or}{
    short = OR,
    long = overlap ratio
}
\DeclareAcronym{snr}{
    short = SNR,
    long = signal to noise ratio
}
\DeclareAcronym{psnr}{
    short = PSNR,
    long = peak signal to noise ratio
}
\DeclareAcronym{kpis}{
    short = KPIs,
    long = key performance indicators
}
\DeclareAcronym{tl}{
    short = TL,
    long = transfer learning
}
\DeclareAcronym{slo}{
  short =  SLO,
  long  = scanning laser ophthalmoscopy
}
\DeclareAcronym{af}{
  short =  AF,
  long  = auto fluorescence
}
\DeclareAcronym{rpe}{
  short =  RPE,
  long  = retinal pigment epithelium
}
\DeclareAcronym{gcl}{
  short =  GCL,
  long  = ganglion cell layer
}
\DeclareAcronym{inl}{
  short =  INL,
  long  = inner plexiform layer
}
\DeclareAcronym{onl}{
  short =  ONL,
  long  = outer nuclear layer
}
\DeclareAcronym{opl}{
  short =  OPL,
  long  = outer plexiform layer
}
\DeclareAcronym{ilm}{
  short =  ILM,
  long  = inner limiting membrane
}
\DeclareAcronym{iz}{
  short =  IZ,
  long  = interdigitation zone
}
\DeclareAcronym{bm}{
  short =  BM,
  long  = bruch membrane
}
\DeclareAcronym{ez}{
  short =  EZ,
  long  = ellipsoidal zone
}
\DeclareAcronym{os}{
  short =  OS,
  long  = outer segment
}
\DeclareAcronym{is}{
  short =  IS,
  long  = inner segment
}
\DeclareAcronym{hf}{
  short =  HF,
  long  = hyper-reflective foci
}

\DeclareAcronym{cslo}{
  short =  cSLO,
  long  = confocal scanning laser ophthalmoscopy
}

\DeclareAcronym{dnn}{
  short = DNN ,
  long  = deep neural networks
}

\DeclareAcronym{pcc}{
  short = PCC ,
  long  = Pearson's correlation coefficient
}

\DeclareAcronym{mri}{
  short = MRI ,
  long  = magnetic resonance imaging
}
\DeclareAcronym{fmri}{
  short = fMRI ,
  long  = functional magnetic resonance imaging
}
\DeclareAcronym{ct}{
  short = CT ,
  long  = computerized tomography
}

\DeclareAcronym{gradcam}{
  short = GradCAM,
  long  = gradient weighted class activation mapping
}

\DeclareAcronym{cam}{
  short = CAM,
  long  = Class activation maps
}
\DeclareAcronym{rcv}{
  short = RCV,
  long  = Regression Concept Vectors
}
\DeclareAcronym{tcav}{
  short = TCAV,
  long  = Testing Concept Activation Vectors
}
\DeclareAcronym{ubs}{
  short = UBS,
  long  = Uniform unit Ball surface Sampling
}
\DeclareAcronym{mls}{
  short = MLS,
  long  = midline shift
}
\DeclareAcronym{gmm}{
  short = GMM,
  long  = Gaussian mixture model
}

\DeclareAcronym{gru}{
  short = GRU,
  long  =  gated recurrent unit
}
\DeclareAcronym{rnn}{
  short = RNN,
  long  =  recurrent neural network
}
\DeclareAcronym{ehr}{
  short = EHR,
  long  =  electronic healthcare record
}

\DeclareAcronym{hilt}{
  short = HITL,
  long  =  human-in-the-loop
}
 
\DeclareAcronym{asd}{
  short = ASD ,
  long  = autism pectrum disorder
}
\DeclareAcronym{mh}{
  short = MH ,
  long  = macular hole
}

\begin{document}
\vspace{-10pt}
 \maketitle

\section*{ABSTRACT}

Deep learning methods for ophthalmic diagnosis have shown considerable success in tasks like segmentation and classification. However, their widespread application is limited due to the models being opaque and vulnerable to making a wrong decision in complicated cases. Explainability methods show the features that a system used to make prediction while uncertainty awareness is the ability of a system to highlight when it is not sure about the decision. This is one of the first studies using uncertainty and explanations for informed clinical decision making. We perform uncertainty analysis of a deep learning model for diagnosis of four retinal diseases - age-related macular degeneration (AMD), central serous retinopathy (CSR), diabetic retinopathy (DR), and macular hole (MH) using images from a publicly available (OCTID) dataset. Monte Carlo (MC) dropout is used at the test time to generate a distribution of parameters and the predictions approximate the predictive posterior of a Bayesian model. A threshold is computed using the distribution and uncertain cases can be referred to the ophthalmologist thus avoiding an erroneous diagnosis. The features learned by the model are visualized using a proven attribution method from a previous study. The effects of uncertainty on model performance and the relationship between uncertainty and explainability are discussed in terms of clinical significance. The uncertainty information along with the heatmaps make the system more trustworthy for use in clinical settings.


\section{Introduction}

Over the last decade, advances in automated detection of diseases using deep learning methods have resulted in performance comparable  to human observers in multiple domains e.g., retinal image diagnosis \cite{de2018clinically, sengupta2020ophthalmic, chap_pending}. The adoption of these methods in clinics is slow if not negligible The key factors hindering trust from end-users, regulators, and patients on deep learning methods in medical imaging are the opaqueness of the algorithms and the tendency to make wrong decisions on complex cases \cite{tomsett2020rapid, singh2020explainable}. There is a need to describe the factors used for making a decision and separating samples with higher uncertainty for expert inspection. The former is the domain of explaining the models and is discussed extensively in the literature \cite{holzinger2017we, arrieta2020explainable, arya2019one}. Studies have been published using a set of explainability methods (called attributions) for retinal image diagnosis \cite{yang2019weakly, sayres2019using, kaur2020interpreting, singh2020interpreting, singh2020optimal}. The attribution methods represent the model output as a sum of the contributions from each input feature. This can be visualized as heatmaps for images and various plots for other data \cite{chen2019explaining}.

The final layer of deep learning models used as classifiers typically consists of softmax outputs corresponding to each of the target classes (usually only one output for a binary classifier). It is often erroneously interpreted as the class probability.  A model can be uncertain in prediction despite having a high softmax value as it learns during training to keep the value at its extremes for a lower loss function. 

Uncertainty is described as the ability of a model to convey the ambiguity in the decision \cite{kendall2017uncertainties}. Deep learning based classifiers typically return the class with the highest softmax output as the result, even if there is a small margin. There is a need for the end-users to be alerted when a model is uncertain to avoid potential wrong decisions. There are two sources of uncertainty in a system - aleatoric and epistemic \cite{tomsett2020rapid, kendall2017uncertainties}. Aleatoric uncertainty or doubt is the inherent uncertainty in the system arising from the modeling process e.g., stochastic behavior. On the other hand, epistemic uncertainty or ambiguity arises due to limited data leading to lower knowledge of the system. The former is inherent and can not be changed for a given model. The latter can be reduced effectively with larger training data and ensuring that new kinds of data (or adversary) are identified. 

The Bayesian theory is used to model the uncertainty of networks but the computational complexity makes it prohibitive for high dimensional problems like image classification. However, existing deep learning models can be cast as Bayesian models without changing the models \cite{gal2016dropout}. The dropout layer which is used only during training is enabled during the testing and the model is run several times for each test sample. This results in an approximate posterior probability distribution for each image. This uncertainty information can improve the diagnostic performance as shown in a study for diagnosing diabetic retinopathy from retinal fundus image \cite{leibig2017leveraging}. It helped identify unfamiliar images that should be referred to a clinician instead of making an ambiguous choice.

A physician can identify when they are uncertain about a case and can access other sources of information (e.g. additional tests, medical history, etc). On the other hand, deep learning methods are have been proposed without specific measures to access their uncertainty. This leads to the challenge of calibrating user trust \cite{tomsett2020rapid} on the system and consequently reduced acceptance.  The samples with high uncertainty can be referred to a human expert to reduce mistakes. Previous studies have indicated that uncertain cases are strongly correlated with errors and identifying them increased model performance e.g. \cite{leibig2017leveraging}. There are some recent studies combining uncertainty and explainability information for 3D object detection \cite{pan2020towards} and clinical time serizes prediction \cite{tan2020explainable, wickstrom2020uncertainty}.

Here, we use a random dropout methodology to approximate a Bayesian neural network to obtain a predictive posterior distribution and use this to quantify uncertainty in diagnosis. We use retinal \ac{oct} images in this study. This uncertainty information along with features (represented as heat maps) used by the model in making the decision is considered. To the best of our knowledge, this is the first study that combines uncertainty with explainability for informed medical diagnosis in addition to being amongst the first studies of its kind in the 2D computer vision domain. 

The custom model with dropout and 6 convolutional layers and the process of finding the uncertainty and explanations are described in section \ref{sec:methods}. The effect of the uncertainty information on the system's performance is studied by adjusting thresholds and removing most uncertain examples in section \ref{sec:results}. Section \ref{sec:discussion} used the explanation heatmaps to discuss the effect of features and input data on model uncertainty. The conclusions and directions for further study are provided in section \ref{sec:conclusion}.

\section{Methods} \label{sec:methods}

\subsection{Dataset}

The OCTID dataset \cite{gholami2020octid} with \ac{oct} images of 4 diseases - \ac{amd}, \ac{csr}, \ac{dr}, and \ac{mh}, as well as normals was used in this study. It provides broad coverage of the most common retinal conditions diagnosed using \ac{oct}. A total of 572 images are distributed unequally over the classes as shown in Table \ref{tab:data}. The normal class is dominant and has about 4 times the number of \ac{amd} scans. The training test split was 80:20 and a further 10\% of the training data was used for validation. After choosing the hyperparameters using a validation set, the model was trained on the entire training set consisting of 459 scans. Data augmentation was performed using rotation, width shift, height shift, shear, and zoom of up to 10\% as well horizontal flip.

\begin{table}[h]
\centering
\caption[Dataset description]{Dataset description showing the class level split for training and test sets.}
\begin{tabular}{| l | l | l | l | l | l | l |}

\hline \textbf{Data} & \textbf{AMD} & \textbf{CSR} & \textbf{DR} & \textbf{MH} & \textbf{Normal} & \textbf{Total} \\ \hline

\textbf{Training} & 44 & 82 & 86 & 82 & 165 & 459\\ \hline
\textbf{Test} & 11 & 20 & 21 & 20 & 41 & 113\\ \hline
\textbf{Total} & 55 & 102 & 107 & 102 & 206 & 572\\ \hline
\textit{\% of total} &  \textit{9.62\%} &  \textit{17.83\%} & \textit{18.71\%} & \textit{17.83\%} &  \textit{36.01\%} & \textit{100\%} \\ \hline

\end{tabular}
 \label{tab:data}
\end{table}

\subsection{Model}

A compact deep learning model with 6 convolutional layers and a dense layer was used. Each block of two convolutional layers of size 16, 32, and 64 was followed by a dropout of 0.2 while the dense layer of 512 neurons was followed by a dropout of 0.3. Initially, it was trained on the UCSD dataset \cite{kermany2018labeled} consisting of 85k images. Since the dataset was large and had low noise the model had very low uncertainty in most cases. Then it was trained on the smaller OCTID dataset which had fewer images and five classes covering multiple retinal diseases simulating a clinical setting. The weights from the model trained on the UCSD dataset were finetuned for OCTID using transfer learning, similar to that used in \cite{singh2019glaucoma}.

\begin{figure}[hbt!]
\captionsetup[subfigure]{position=b}
\centering
\begin{subfigure}{.5\textwidth}
  \centering
  \includegraphics[width=.99\linewidth]{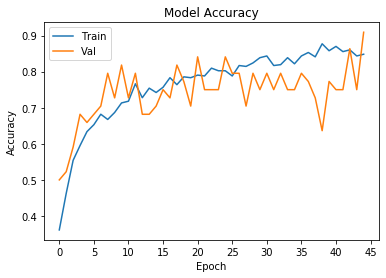}
  \label{fig:acc}
\end{subfigure}%
\begin{subfigure}{.4\textwidth}
  \centering
  \includegraphics[width=.99\linewidth]{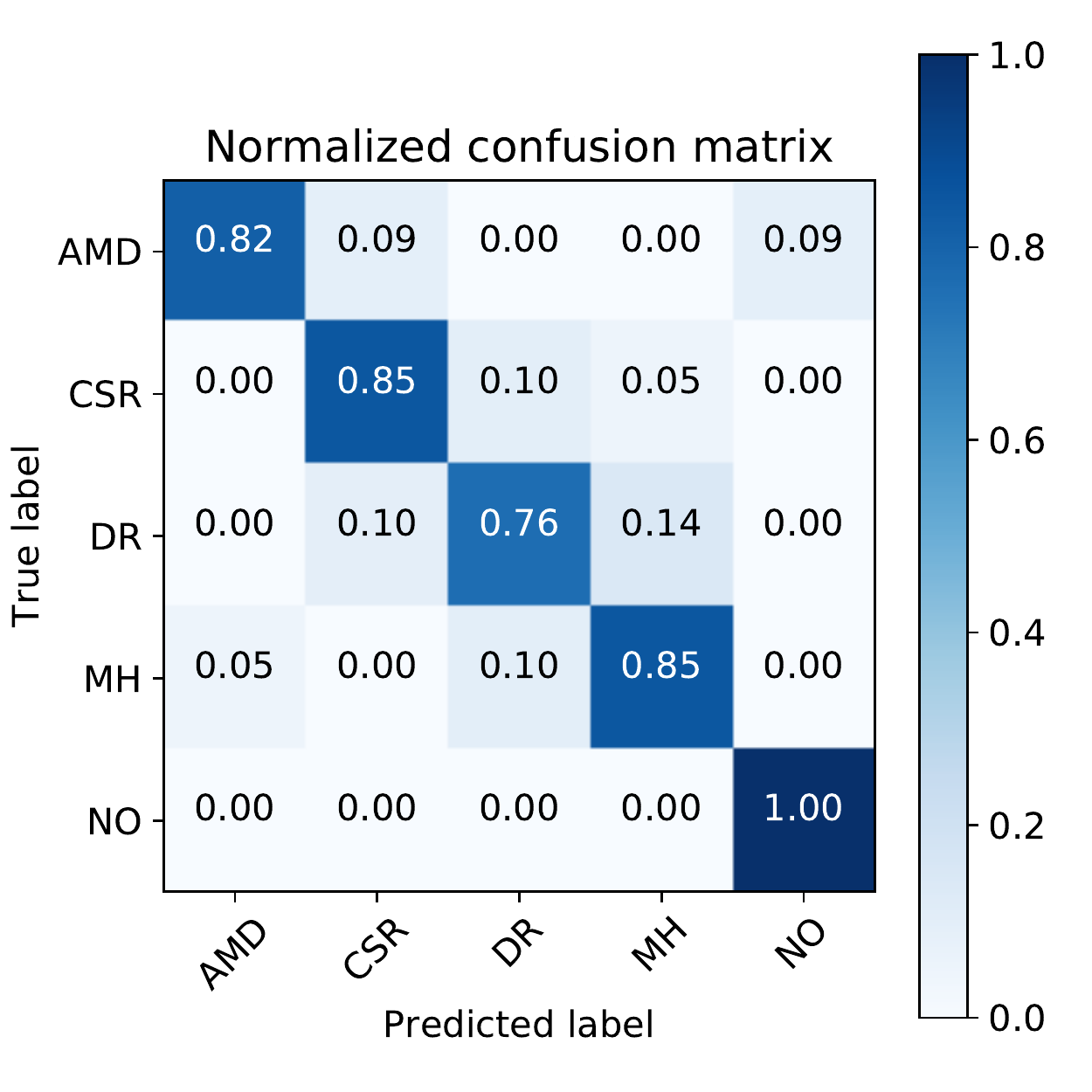}
  \label{fig:conf}
\end{subfigure}
\caption{The metrics of the original model - (left) evolution of training and validation accuracy during training, (right) confusion matrix for the test set}
\label{fig:org}
\end{figure}

The training progress over 45 epochs in Figure \ref{fig:org} (left) shows a final training accuracy of 84.8\% and a validation accuracy of 90.9\%. The validation accuracy graph is bumpy due to small number of images. The resulting model obtained a test accuracy of 88.5\% and the confusion matrix is shown in Figure \ref{fig:org} (right). It shows that the model had the weakest performance for \ac{dr}, possibly because of large amount of variation among the images. All the normal images were correctly predicted but 9\% of the \ac{amd} images were categorized as normal.

\subsection{Uncertainty evaluation and explanation}

Some of the model weights were removed randomly using the 4 dropout layers at the test time to approximate a Bayesian neural network. The 10\textsuperscript{th} percentile level from the training set was used to generate thresholds for each class in the initial analysis. The test samples with uncertainty higher than the threshold (lower probability of prediction) were marked as referrals for manual diagnosis by an expert.

The explanations were produced using an attribution method called DeepTaylor \cite{montavon2017explaining}. It was the best performing attribution method from a previous comparative study \cite{singh2020optimal}. It finds a root point near a given neuron with a value close to the input but output as 0. Taylor decomposition is then recursively applied from the output to the input layer to estimate the attributions of each neuron. The Innvestigate library \cite{alber2019innvestigate} was used to implement the attribution method.

\section{Results} \label{sec:results}

\begin{figure}[h!] 
\centering
\includegraphics[width=0.4\linewidth]{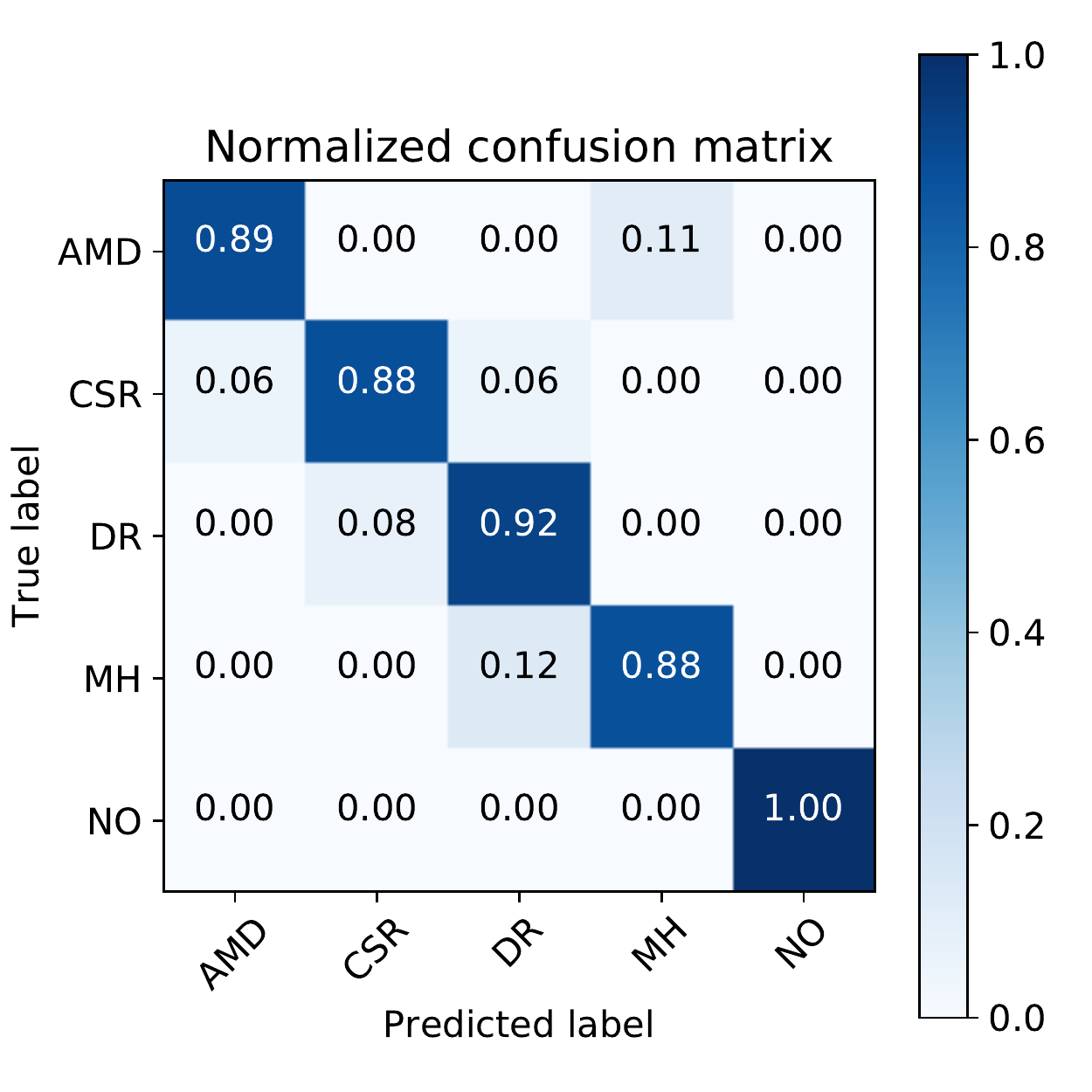}
\caption{Confusion matrix for the test set after uncertainty threshold} \label{fig:conf-new}
\end{figure}

A threshold was computed using the 10th percentile of the uncertainty in the training set for each class. These values varied across the classes - \ac{amd} 0.65, \ac{csr} 0.92, \ac{dr} 0.93, \ac{mh} 0.97, and normal 0.93. The lower threshold for \ac{amd} indicates a wider spread in uncertainty levels and the presence of false negatives. The thresholds were used to remove 27 of 113 test images and the model accuracy improved from 88.5\% to 93.7\% for the remaining samples. These improvements for each class are shown by metrics in Table \ref{tab:metrics} and the new confusion matrix is shown Figure \ref{fig:conf-new}. 

\ac{dr}, which had the worst performance originally as shown in Figure \ref{fig:org} (right), showed the largest improvement from 0.76 to 0.92 at the cost of separating 6/21 images for referral. In contrast, only one normal image was sent for referral and zero false positives were maintained. Interestingly, all the false negatives were also eliminated by referring 4/13 \ac{amd} images. The fraction of correct classifications also increased for the other three classes - 0.82 to 0.89 for \ac{amd}, 0.85 to 0.88 for \ac{csr}, and 0.88 for \ac{mh}. It is apparent from this case that the largest improvements were made in the classes with weakest performance, but no conclusions can be reached without an analysis involving larger and more diverse datasets. 

\begin{table}[h]
\begin{center}

\caption{Effect of threshold on precision (PR), recall (RE), F1 score and support samples}\label{tab:metrics}
\begin{tabular}{|c | c c c c | c c c c|}
\cline{1-9}
 & \multicolumn{4}{c|}{\textbf{Base metrics}} & \multicolumn{4}{c|}{\textbf{After uncertainty}}\\
\cline{2-9}
\textbf{Condition}& \textbf{PR} & \textbf{RE} & \textbf{F1} & \textbf{Support} & \textbf{PR} & \textbf{RE} & \textbf{F1} & \textbf{Support} \\
\hline
AMD & 0.90 & 0.82 & 0.86 & 11 & 0.89 & 0.89 & 0.89 & 9 \\ \hline
CSR & 0.85 & 0.85 & 0.85 & 20 & 0.88 & 0.94 & 0.91 & 16\\ \hline
DR & 0.80 & 0.76 & 0.78 &  21 & 0.92 & 0.80 & 0.86 & 15\\ \hline
MH & 0.81 & 0.85 & 0.83 & 20 & 0.88 & 0.93 & 0.90 & 15 \\ \hline
Normal & 0.95 & 0.95 & 0.95 & 41 & 1.00 & 1.00 & 1.00 & 40 \\ \hline

\end{tabular}
\end{center}
\end{table}

 \begin{figure}[hbt!] 
\centering
\includegraphics[width=0.6\linewidth]{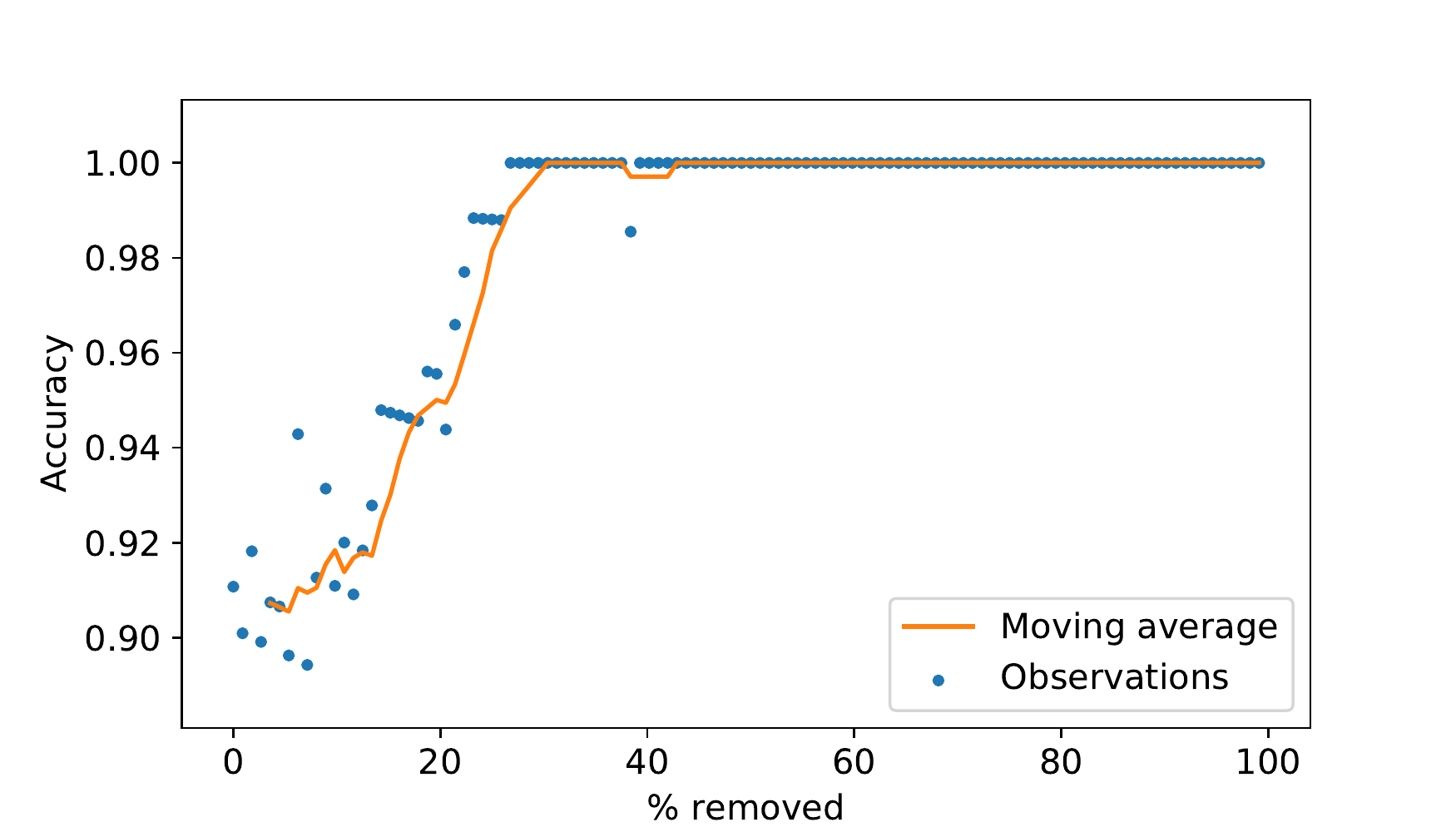}
\caption{The effect of removing the samples with most uncertainty on the accuracy. The points in the scatter are the observations and the 5 sample moving average shows the overall trend.} \label{fig:moving-avg}

\end{figure}

Probability distributions of the output were generated by activating random dropout at the test time and repeating for 1000 times for the entire dataset. These probability distribution histograms approximating the Bayesian uncertainty and the corresponding \ac{oct} images with heatmaps using Deep Taylor are shown for some typical samples in Figure \ref{fig:vis}. The brighter the color on a given part of the image, the greater the contribution to the model output. The original \ac{oct} images are also provided for reference. The threshold is plotted as a solid green and the medians of each class are shown by dotted lines. If the class median is below the threshold level, the image should be sent for a referral. The observed relationship between the explanations, uncertainty levels and the model predictions is discussed in section \ref{sec:discussion}.

An analysis performed by removing the test samples with most uncertainty and observing the effect on the model accuracy is plotted in Figure \ref{fig:moving-avg}. The samples were ordered by the median value of the  uncertainty and removed one by one. Small sample size of 113 makes the overall trend noisy and hence moving average analysis with window size 5 was used for smoothing. Almost all the incorrectly classified images were in the initial quarter of uncertainty scores. The moving average has a shape similar to the logistic growth as it reaches 100\% accuracy level. In practice, such a graph could be used to set a tolerance level and images below the level can be referred to clinician for diagnosis by a human expert. In further studies with larger datasets, this could be used to produce class wise thresholds from the test set.







\section{Discussion} \label{sec:discussion}

The black-box nature of deep learning models affects their acceptance in practical settings such as healthcare. A clinician is supposed to present the reasoning for decisions. Humans also associate a degree of certainty to their decisions. This is in contrast to the opaqueness of advanced artificial intelligence methods such as deep learning which are hard to understand due to millions of features and give the result as a single class. This study presented an uncertainty aware and explainable deep learning model for retinal diagnosis. The heatmaps using Deep Taylor and the uncertainty using probability histograms supplement the model decision. The end-user can infer the regions the model looked at as well as how sure the model is about the predictions and make a final decision. Also, separating the cases with more uncertainty for referral could help improve model performance and inculcate trust. The improved confusion matrix and accuracy in Figures \ref{fig:conf-new} and \ref{fig:moving-avg} respectively indicate the efficacy of uncertainty information. Referring uncertain images to a clinician instead of misdiagnosing can improve patient care by increasing the confidence in the underlying system.

 Previous studies \cite{singh2020interpreting, singh2020optimal} compared the heatmaps of each explainability method for a given sample and concluded that Deep Taylor performed the best for retinal \ac{oct}. There are also studies showing the effect of uncertainty for diagnosis from retinal fundus images \cite{leibig2017leveraging}. The present study is one of the first to report both the explanations and uncertainty analysis. It is observed that the samples classified with higher certainty also have more relevant heatmaps. Some examples covering all 5 classes and some typical cases are shown in Figure \ref{fig:vis}. A greater emphasis is laid on the ability to identify model errors using uncertainty and explanations. Figure \ref{fig:normal-corr} shows the retinal \ac{oct} of a normal eye for reference.
 
 The model is observed to be the most sensitive to high contrast regions, i.e. the inner retina or the vitreous - retinal interface and the photoreceptor layer. Figure \ref{fig:amd-corr} shows a correctly classified \ac{amd} scan where the model showed certainty higher than the corresponding threshold. The model correctly picks up the relevant deposits and new blood vessels in the photoreceptor layer. Figures \ref{fig:csr-corr} - \ref{fig:mh-corr} show correctly classified CSR, DR and MH cases.
 
 \newpage
 
 \begin{figure}[hb!]

\begin{subfigure}{\textwidth}
\centering
\begin{subfigure}{.62\textwidth}
  \centering
  \includegraphics[width=.99\linewidth]{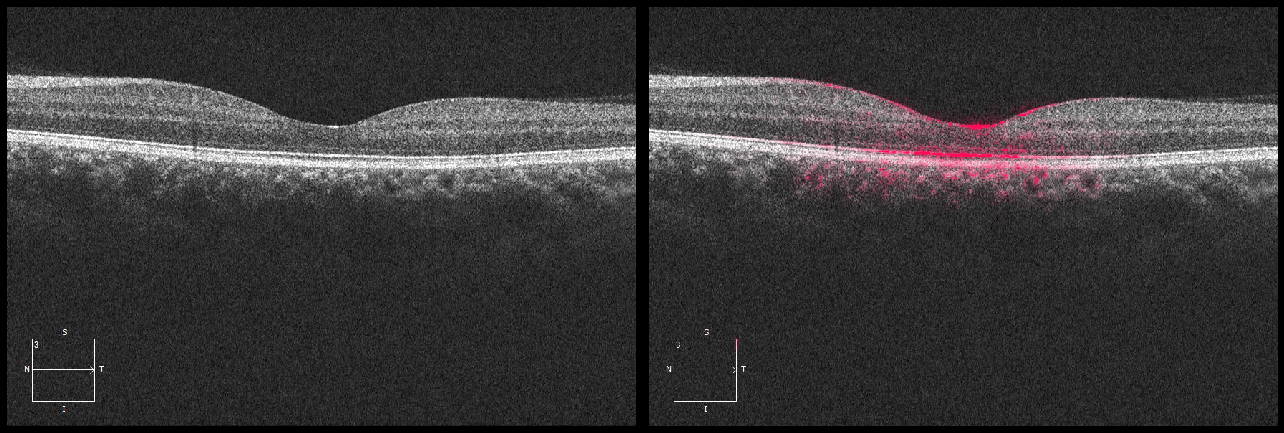}
\end{subfigure}%
\begin{subfigure}{.38\textwidth}
  \centering
  \includegraphics[width=.99\linewidth]{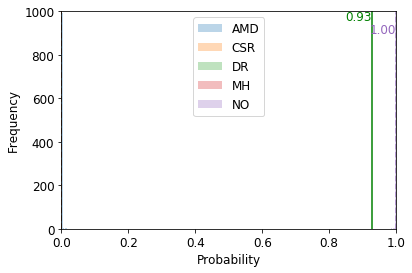}
\end{subfigure}
\subcaption{A correctly classified normal image with high certainty. The probability was close to 1 for all the runs. The heatmaps focus on the normal shapes of inner retina and the photoreceptor layer.}
\label{fig:normal-corr}
\end{subfigure}

\begin{subfigure}{\textwidth}
\centering
\begin{subfigure}{.62\textwidth}
  \centering
  \includegraphics[width=.99\linewidth]{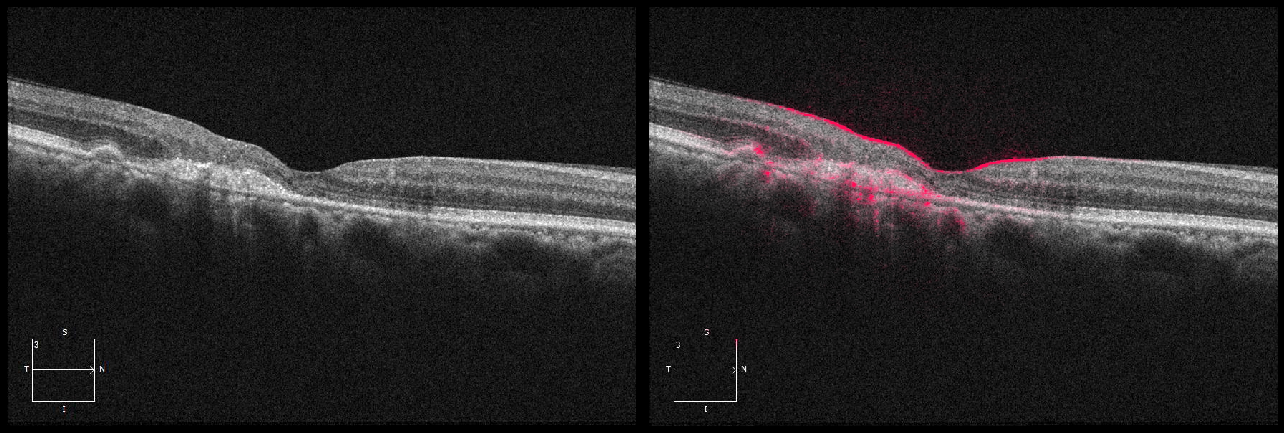}
\end{subfigure}%
\begin{subfigure}{.38\textwidth}
  \centering
  \includegraphics[width=.99\linewidth]{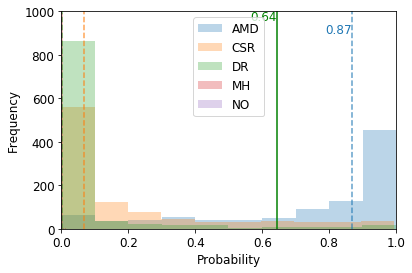}
\end{subfigure}
\subcaption{An \ac{amd} image correctly classified with probability higher than the threshold. The explanations highlight the anomalous photoreceptor layer in the macula.}
\label{fig:amd-corr}
\end{subfigure}

\begin{subfigure}{\textwidth}
\centering
\begin{subfigure}{.62\textwidth}
  \centering
  \includegraphics[width=.99\linewidth]{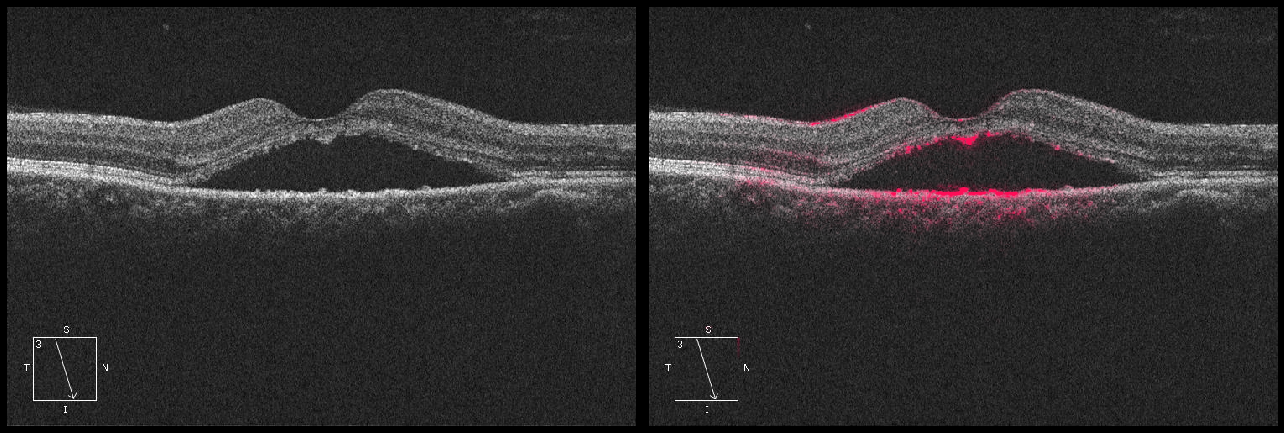}
\end{subfigure}%
\begin{subfigure}{.38\textwidth}
  \centering
  \includegraphics[width=.99\linewidth]{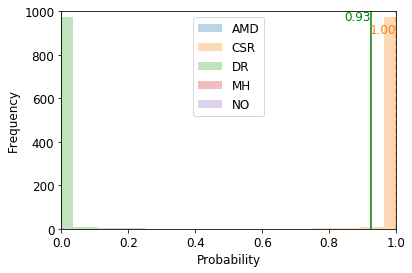}
\end{subfigure}
\subcaption{A \ac{csr} image with prominent features correctly classified with complete certainty.}
\label{fig:csr-corr}
\end{subfigure}

\begin{subfigure}{\textwidth}
\centering
\begin{subfigure}{.62\textwidth}
  \centering
  \includegraphics[width=.99\linewidth]{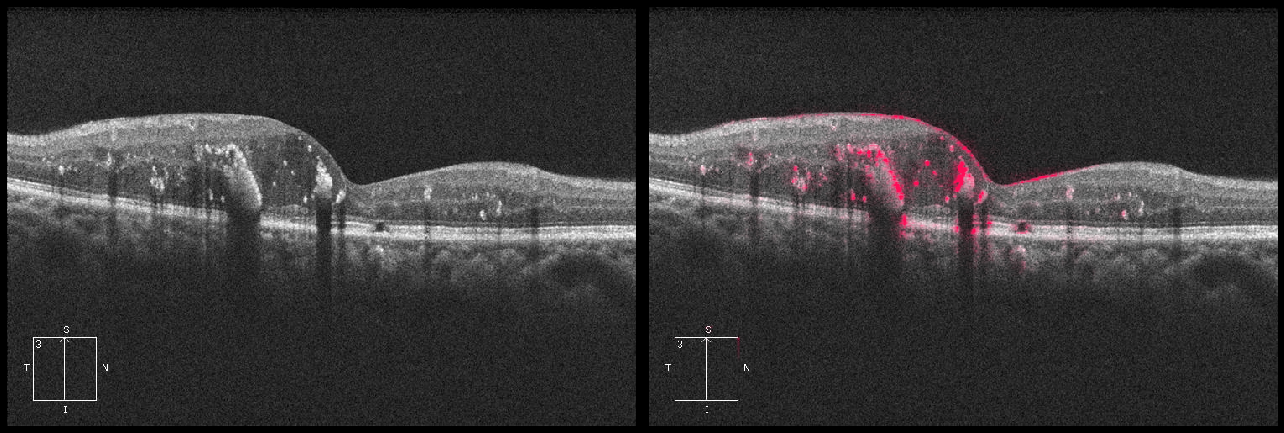}
\end{subfigure}%
\begin{subfigure}{.38\textwidth}
  \centering
  \includegraphics[width=.99\linewidth]{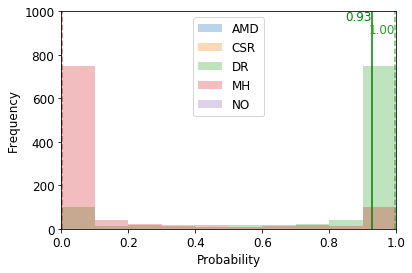}
\end{subfigure}
\subcaption{A correctly classified \ac{dr} image with high certainty. The model detected new blood vessels visible as bright structures with shadows underneath.}
\label{fig:dr-corr}
\end{subfigure}

\end{figure}

\newpage

\begin{figure} \ContinuedFloat
\centering

\begin{subfigure}{\textwidth} 
\centering
\begin{subfigure}{.62\textwidth}
  \centering
  \includegraphics[width=.99\linewidth]{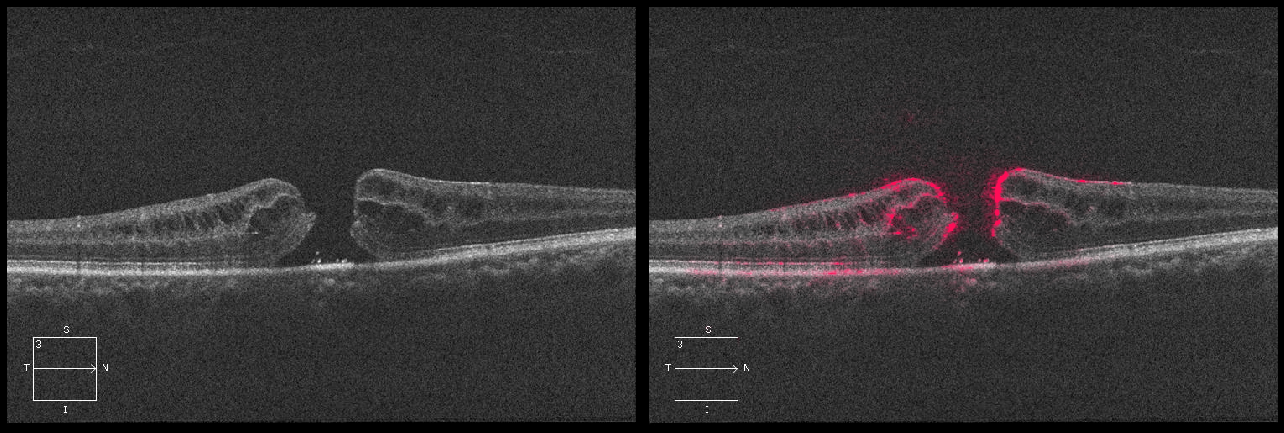}
\end{subfigure}%
\begin{subfigure}{.38\textwidth}
  \centering
  \includegraphics[width=.99\linewidth]{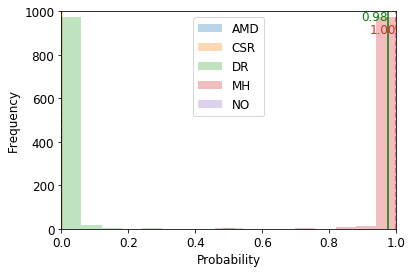}
\end{subfigure}
\subcaption{A correctly classified \ac{mh} image. The certainty is high and the model looked the curved boundaries of the hole.}
\label{fig:mh-corr}
\end{subfigure}


\begin{subfigure}{\textwidth}
\centering
\begin{subfigure}{.62\textwidth}
  \centering
  \includegraphics[width=.99\linewidth]{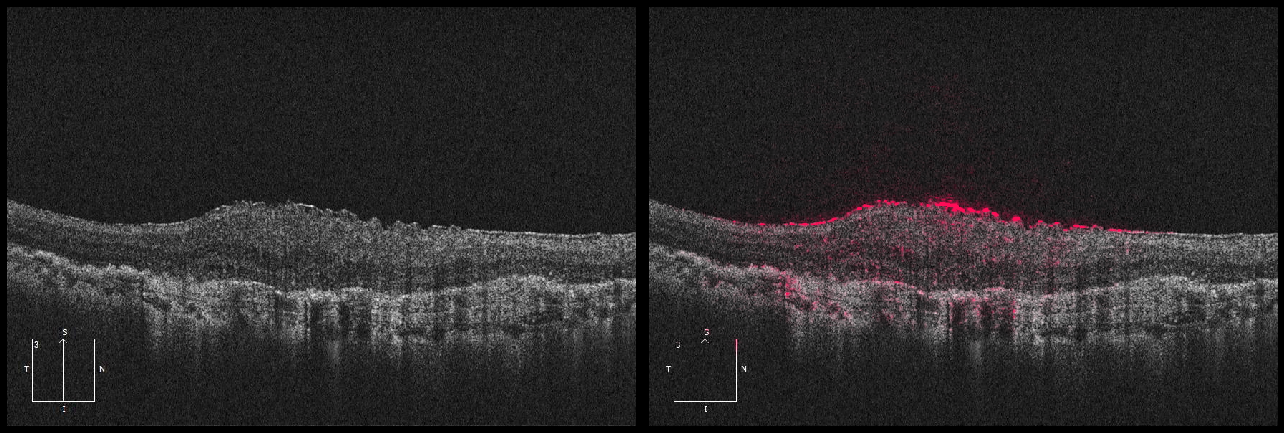}
\end{subfigure}%
\begin{subfigure}{.38\textwidth}
  \centering
  \includegraphics[width=.99\linewidth]{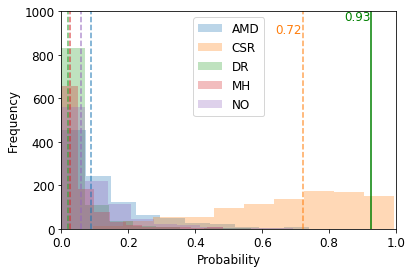}
\end{subfigure}
\subcaption{A noisy \ac{amd} image misclassified as \ac{csr} with certainty lower than the threshold. Hence, model refers it to a clinician for further diagnosis.}
\label{fig:amd-ref}
\end{subfigure}

\begin{subfigure}{\textwidth}
\centering
\begin{subfigure}{.62\textwidth}
  \centering
  \includegraphics[width=.99\linewidth]{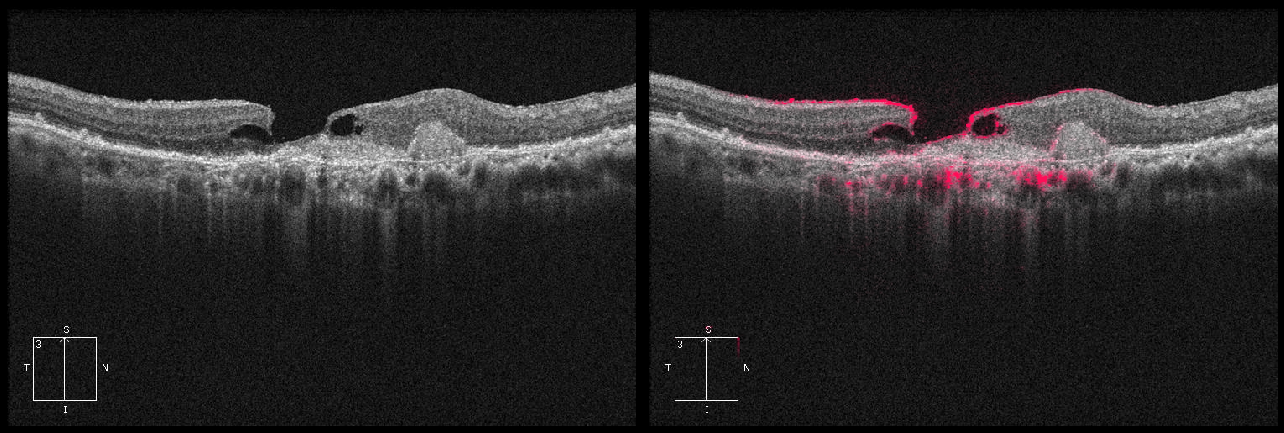}
\end{subfigure}%
\begin{subfigure}{.38\textwidth}
  \centering
  \includegraphics[width=.99\linewidth]{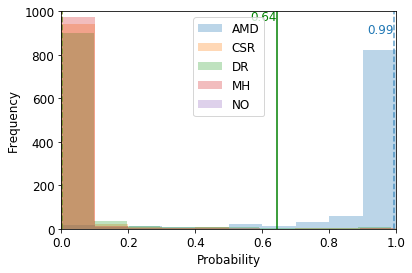}
\end{subfigure}
\subcaption{An image labelled as \ac{mh} in the dataset with both \ac{mh} and \ac{amd} classified as \ac{amd} with high certainty.}
\label{fig:mh-amd-sec}
\end{subfigure}

\vspace{10pt}

\caption{(Left to right) The input OCT images, explanations using Deep Taylor and Bayesian uncertainty histograms. The brighter magenta regions had more effect on the output. The histograms shows the distribution of softmax outputs for 1000 runs of the model with medians as dotted and threshold in solid green.} \label{fig:vis}
\end{figure}

\vspace{-10pt}
 
 A typical example of \ac{csr} is shown in Figure \ref{fig:csr-corr} with as large fluid deposit in the central part of the eye. The model looked at the boundaries of the deposit and correctly classified it with high certainty. Similarly, Figure \ref{fig:dr-corr} shows a \ac{dr} which is also classified correctly with high certainty. The model looked at the irregular structures, possibly blood clots or scars in the central retina. The abnormalities in the periphery are not highlighted in most cases. A correctly classified and highly certain \ac{mh} case is shown in Figure \ref{fig:mh-corr} where the model mainly looked at the main clinical feature - the boundaries of the hole. Figure \ref{fig:amd-ref} is a high noise image with diffuse photoreceptor layer. The model emphasized the curvature of the retina while focusing less on the telltale fluid deposits whose boundaries are highlighted in a light magenta. Consequently, \ac{mh} received the highest probability. The higher threshold of \ac{mh} ensured the case is sent for a referral instead of misdiagnosing it. Figure \ref{fig:mh-amd-sec} shows an image which has a clear \ac{mh} and a possible wet \ac{amd}. It was labeled as \ac{mh} in the dataset and the model classified it as \ac{amd}. It put emphasis on the macular deposits compared to the relatively smaller structure of the hole, thus predicting the secondary diagnosis.
 
 Some general observations can also be drawn about the relationship between uncertainty and the features highlighted by explanations. The cases with higher contrast between the structures seem to be classified with higher certainty. The model looked at the sharp transitions between layers such as the vitreous - retinal and the photoreceptors along with some bright deposits as in Figure \ref{fig:dr-corr}. Clinicians also consider the darker regions such as the fluid accumulations in Figure \ref{fig:csr-corr} and the shadows of new blood vessels in Figure \ref{fig:dr-corr}. Overall, there is a relationship between the uncertainty, explanations, and the accuracy of a deep learning model for retinal \ac{oct} images. This study is an initial exploratory analysis and leads to several areas of further research.

\section{Conclusion} \label{sec:conclusion}

In this study, the Bayesian uncertainty and the explanations of a  deep learning model for retinal diagnosis are demonstrated. It is shown that the threshold with uncertainty successfully improved the model performance by leaving out a few uncertain samples. The effect of removing uncertain samples on the accuracy follows a trend similar to the logistic curve. The uncertainty and the correctness of a decision are related to the explanations. In cases with higher certainty and correct decisions the explanations highlighted the clinically relevant regions the best.  Using both uncertainty and explainability has implications in the acceptance of deep learning models. Both the clinical community and AI researchers stand to benefit from this, the former with more holistic information and the latter with a tool to improve models. 

Future studies can expand this approach to more domains for improving the deep learning models for real-world applications. This information can be used to alter the features learned by a model and hence improving the robustness. Another direction could be to quantify pathological features such as the brightness and thickness of the retinal pigment epithelium and relate them to the explanations and uncertainty. Designing an uncertainty-aware and explainable system would alleviate the major barriers for acceptance of deep learning methods in multiple domains including medical diagnosis.

\section*{Acknowledgement}
This work is supported by an NSERC Discovery Grant and NVIDIA Titan V GPU Grant to V.L. This research was enabled in part by Compute Canada (www.computecanada.ca).
\clearpage
\bibliography{main.bib} 
\bibliographystyle{spiebib} 
\end{document}